\documentclass[12pt]{iopart}

\usepackage{amssymb}
\usepackage{graphicx}
\usepackage{bm}
\usepackage{verbatim}
\usepackage{tensor}
\usepackage{undertilde}
\usepackage{epstopdf}
\usepackage{cite}

\DeclareMathAlphabet{\mathpzc}{OT1}{pzc}{m}{it}

\begin{document}


\title[]{Scalar QNMs for higher dimensional black holes surrounded by quintessence in Rastall gravity}

\author{J. P. Morais Gra\c ca$^{1}$ and Iarley P. Lobo$^{1}$}

\address{$^{1}$ Departamento de F\'{i}sica, Universidade Federal da Para\'{i}ba, Caixa Postal 5008, CEP 58051-970, Jo\~{a}o Pessoa, PB, Brazil}

\ead{jpmorais@gmail.com and iarley\_lobo@fisica.ufpb.br}

\begin{abstract}
The spacetime solution for a black hole, surrounded by an exotic matter field, in Rastall gravity, is calculated in an arbitrary d-dimensional spacetime. After this, we calculate the scalar quasinormal modes of such solution, and study the shift on the modes caused by the modification of the theory of gravity, i.e., by the introduction of a new term due to Rastall. We conclude that the shift strongly depends on the kind of exotic field one is studying, but for a low density matter that supposedly pervades the universe, it is unlikely that Rastall gravity will cause any instability on the probe field.
\end{abstract}




\maketitle


%
%
\section{Introduction}

Rastall gravity is inspired by the idea that conservation laws based on spacetime symmetries has been probed only in the flat or weak field regime of gravity \cite{Rastall:1973nw}. Within this approach, the covariant derivate of the energy-momentum tensor does not need to vanish, but can be a function of the derivative of some scalar curvature. One interesting result of this model is that all electrovacuum solutions of general relativity are also solutions of Rastall gravity, but as long as (non-vanishing trace) matter is introduced in the theory, the spacetime becomes dependent of the Rastall parameter.

This framework has been attracted some attentions over the last few years, and some results have been already obtained, both related to astrophysical \cite{deMello:2014hra,Oliveira:2015lka,Bronnikov:2016odv,vel,Heydarzade:2016zof,Heydarzade:2017wxu,Licata,gaus} and cosmological solutions \cite{Capone:2009xm,Batista:2011nu,Silva:2012gn,Santos:2014ewa,hm,Yuan:2016pkz,Lobo:2017dib}. In particular, black holes solutions surrounded by exotic matter fields have been obtained based on the idea developed by Kiselev  \cite{Kiselev:2002dx}, who states that one should consider astrophysical solutions surrounded by a dark energy field, in accordance with the recent cosmological observations about the acceleration of the universe.

In their original work, Kiselev considered black holes surrounded by a quintessence field, but the approach used by the author allows for any kind of isotropic fluid with a defined equation of state. In the following, Chen et al \cite{Chen:2008ra} found the solution for higher dimensional black holes surrounded by quintessence, in general relativity, and Heydarzade e Darabi developed the analogue four-dimensional solution in the framework of Rastall gravity. Our first goal in this paper is to use the same approach to found the higher dimensional analogue of \cite{Chen:2008ra} in Rastall gravity.

Also in this paper, our aim is to study the stability of scalar probe fields on the spacetime generated by higher dimensional black holes surrounded by exotic matter fields in Rastall gravity. To be more precise, we will study the scalar quasinormal modes for these black holes. Quasinormal modes are complex numbers that model the emission of gravitational waves by perturbed objects. Its real part is related to the frequency of emission, and its imaginary part is related to its damping (For a review, see \cite{Berti:2009kk}).

The study of quasinormal modes has received a great deal of attention over the last years due to both experimental and theoretical reasons: The most obvious one is related to the detection of gravitational waves emitted by compact objects, such as black holes and neutron stars \cite{Abbott:2016blz}. Also as a theoretical tool, we can mention the so-called gauge/gravity dualities, where the quasinormal modes is related to the poles of a propagator in the dual field theory. Being just a mere theoretical framework for the study of strong coupled gauge theories, or a realistic description of a hypothetical holographic universe, the gauge/gravity dualities has attained a great deal of attention over the last years.

The quasinormal modes for black holes surrounded by quintessence in a four-dimensional spacetime governed by general relativity has already been calculated in \cite{Chen:2005qh,Zhang:2006ij,Zhang:2006hh,Ma:2006by,Zhang:2007nu}. We will perform the calculations in the framework of Rastall gravity, for higher dimensional black holes. 

This work is organized as following. In chapter 2, we will introduce the framework of Rastall gravity, and calculate the spacetime generated by a black hole surrounded by a homogeneous and isotropic matter field, for an arbitrary higher dimension. In section three we will briefly present the WKB approach for the calculation of the quasinormal modes, and calculate the quasinormal modes for both a quintessence and a phantom field. Finally, in chapter 4 we will present our conclusions.

\section{Higher dimensional BHs surrounded by a perfect fluid}

The idea behind Rastall model is that the well-known conservation laws have been tested only in the weak field regime of gravity. Adopting this perspective, the covariant derivative of the energy-momentum tensor does not need to vanish, instead it is proportional to the derivative of some curvature invariant. A straightforward choice is to consider that

\begin{equation}
\nabla_{\mu} T^{\mu\nu} \propto \nabla^\mu R,
\end{equation}
where $R$ is the Ricci scalar. This leads to a modification of Einstein's equations, that can be recast as

\begin{equation}
G_{\mu\nu} + \kappa \lambda g_{\mu\nu} R = \kappa T_{\mu\nu},
\label{RastallEq}
\end{equation} 
where $\kappa$ is a constant related to the Newton's gravitational constant. Taking the trace of Eq. (\ref{RastallEq}), one gets

\begin{equation}
R(4\kappa\lambda-1) = \kappa T.
\end{equation}

For a vanishing trace of the energy-momentum tensor, such as the electrovacuum solution, the above equation requires that $\kappa\lambda = 1/4$ or $R=0$. The former possibility is not physically acceptable, since it would demands that $T=0$ for any scalar curvature. This lead us to conclude that for all matter configuration where the energy-momentum tensor has null trace, Rastall theory have the same solutions as general relativity.

This feature of Rastall model has lead some authors to search for black holes solutions in a background of matter/energy where its energy-momentum tensor has non-vanishing trace. This is a physically acceptable procedure, since the most acceptable explanation for the current acceleration of the universe is the presence of a kind of exotic dark energy, that pervades all the spacetime.

In this paper we will consider two possibilities for this dark energy, to known, a quintessence field and a phantom field. If one adopts a Schwarzschild-like line element, 

\begin{equation}
ds^2 = f(r) dt^2 - f(r)^{-1} dr^2 - r^2 d \Omega_{d-2},
\label{metric}
\end{equation}
where $d\Omega_{d-2}$ stands for the $(d-2)$-dimensional sphere, the energy-momentum tensor for a homogeneous and isotropic matter field can be written as \cite{Kiselev:2002dx,Chen:2008ra}

\begin{equation}
T\indices{^t_t} = T\indices{^r_r} = \rho(r)
\end{equation}
and
\begin{equation}
T\indices{^{\theta_1}_{\theta_1}} = T\indices{^{\theta_2}_{\theta_2}} = ... = T\indices{^{\theta_{d-2}}_{\theta_{d-2}}} = - \frac{1}{d-2} \rho(r)[(d-1)\omega+1],
\end{equation}
where $\omega$ is the parameter for the equation of state of the kind of field we are considering.

Our aim in this section is to solve the gravitational field equations, Eq. (\ref{RastallEq}), for the above energy-momentum tensor, considering the line element given by Eq. (\ref{metric}). Considering the spherical symmetry of the problem, we have only two independent field equations,

\begin{eqnarray}
\nonumber
\frac{1}{2r^2}[f'r(d-2)-(d-3)(d-2)(1-f)-\kappa\lambda(2f''r^2 
\\
+4(d-2)f'r-2(d-3)(d-2)(1-f))] = \kappa T\indices{^t_t}
\label{eq1}
\end{eqnarray}
and
\begin{eqnarray}
\nonumber
\frac{1}{2r^2}[f''r^2+2(d-3)f'r-(d-4)(d-3)(1-f)-
\\2\kappa\lambda(f''r^2+2(d-2)f'r-(d-3)(d-2)(1-f))]=\kappa T\indices{^{\theta_d}_{\theta_d}},
\label{eq2}
\end{eqnarray}
and two independent functions, $f(r)$ and $\rho(r)$. This system can be solved imposing a power law function for the density, $\rho(r) =  Nr^A$. For the consistency between equations (\ref{eq1}) and (\ref{eq2}), one must have that

\begin{equation}
f(r) = 1 + \frac{C_1}{r^{d-3}} - N_s r^{-\frac{(d-2)(d-1)(1+\omega)-2\kappa\lambda d(d-1)(1+\omega)}{(d-2)-2\kappa\lambda (d-1)(1+\omega)}+2},
\label{fmetric}
\end{equation}
and 
\begin{equation}
\rho(r) = - \frac{1}{2}\frac{\mathcal{W}N_s}{\kappa}r^{-\frac{(d-2)(d-1)(1+\omega)-2\kappa\lambda d(d-1)(1+\omega)}{(d-2)-2\kappa\lambda (d-1)(1+\omega)}},
\end{equation} 
where

\begin{equation}
\mathcal{W} = \frac{[ \omega(d-2)(d-1)-2\kappa\lambda (d-1)(1+\omega)](2\kappa\lambda d-d+2)(d-2)}{[(d-2)-2\kappa\lambda(d-1)(1+\omega)]^2},
\end{equation}
and $C_1$ and $N_s$ are constants. The former is related to the mass of the black hole, and the latter related to the density of the exotic matter. In the case where $d=4$, the above solution reproduces \cite{Heydarzade:2017wxu}, and in the case where $\kappa \lambda \rightarrow 0$, the above results reproduce \cite{Chen:2008ra}.

\subsection{Horizons}

To find the horizons of the metric (\ref{metric}), for which the $g_{00}$ coefficient is given by the function (\ref{fmetric}), one must determine the zeros of such function. In general, this can be achieved only by using numerical methods. 

For most of our chosen configurations, the metric will resemble a Scharzschild-de Sitter spacetime, i.e., will have an event horizon and a cosmological horizon, and for our purpose one must consider only the spacetime region between these radial coordinates.

\section{Scalar quasinormal modes}
The most straightforward method to study perturbations near a spacetime generated by a black hole is to allow probe fields to be perturbed by such spacetime, without backreacting on it. In general, for a scalar field, this means to find solutions for the Klein-Gordon equation for a well-defined boundary condition, to known, the condition that waves go into the event horizon and also to radial infinity.

In Rastall gravity, it is not possible to determine the exact field equation a scalar field should obey, since to the present moment there is no well accepted Lagrangian for the theory, but one can guess that the Klein-Gordon equation is a first approximation to the modified equation. Any correction to the field equation due to some non-minimal coupling with the curvature can be ignored in such first approximation. Then, the corrections on the quasinormal modes due to the additional Rastall coupling will appear only on the metric.

The dynamics of a massless scalar field $\Phi(\textbf{x})$, in a $d$-dimensional spacetime, is then given by the equation $\nabla_\mu \nabla^\mu \Phi(\textbf{x}) = 0$. For the static spherically symmetric line element defined by Eq. (\ref{metric}), the field equation can be reduced to a Schr\"odinger type equation, namely,

\begin{equation}
\left( \frac{d^2}{dr^{*2}} + \omega^2 - V(r) \right) \Phi(\textbf{r}) = 0,
\label{scalarEquation}
\end{equation}
where the tortoise coordinate is defined as $dr/dr^* = f(r)$,  and the effective potential $V(r)$ is given by

\begin{equation}
V(r) = f(r) \left(\frac{(d-2)(d-4)}{4r^2}f(r) + \frac{d-2}{2r}f'(r) + \frac{l(l+D-3)}{r^2}\right),
\end{equation}
with $l$ being the angular momentum eigenvalue related to the angular momentum operator $L^2$. 

\begin{figure}[]
\centering
\begin{tabular}{@{}cc@{}}
\includegraphics[scale=0.4]{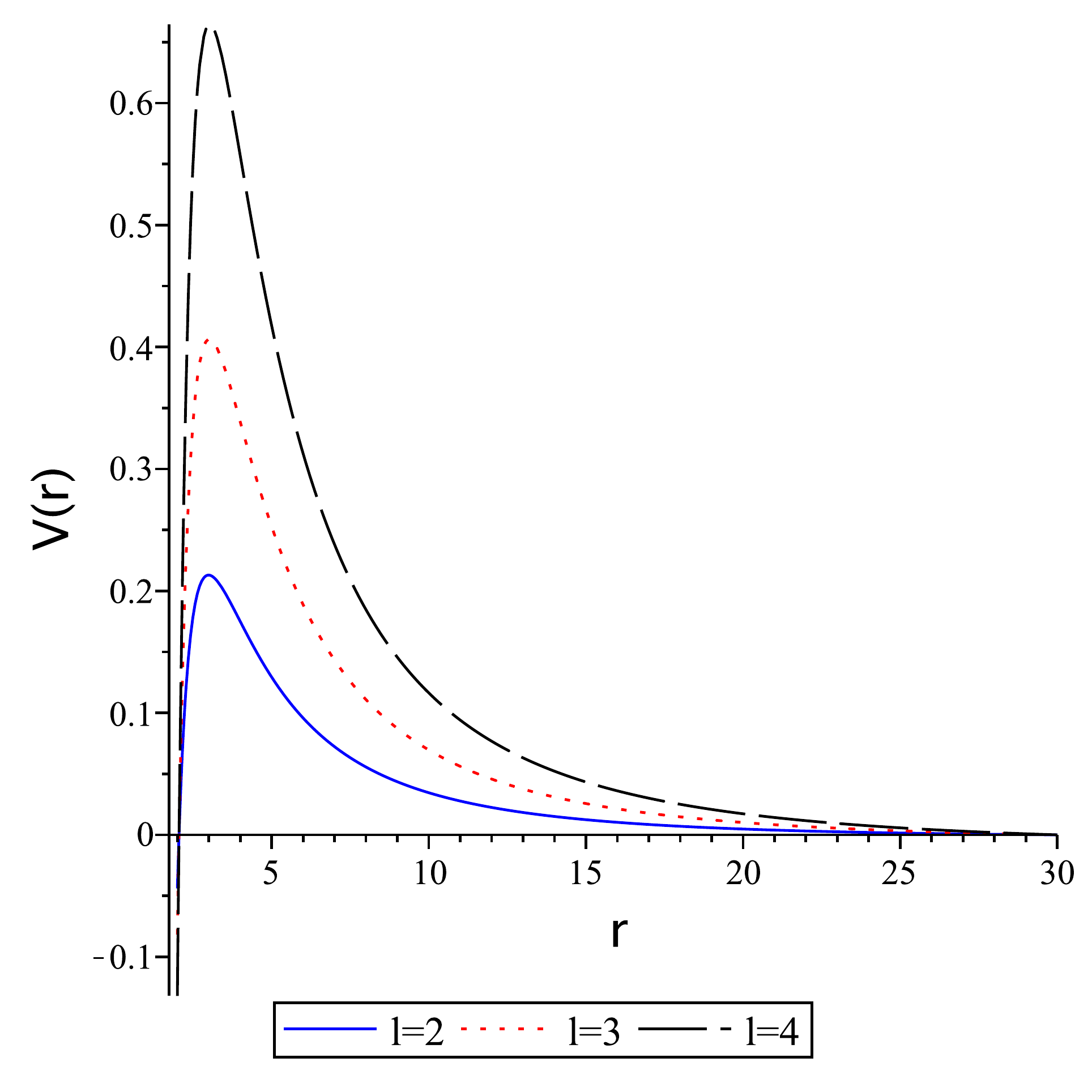} &
\includegraphics[scale=0.4]{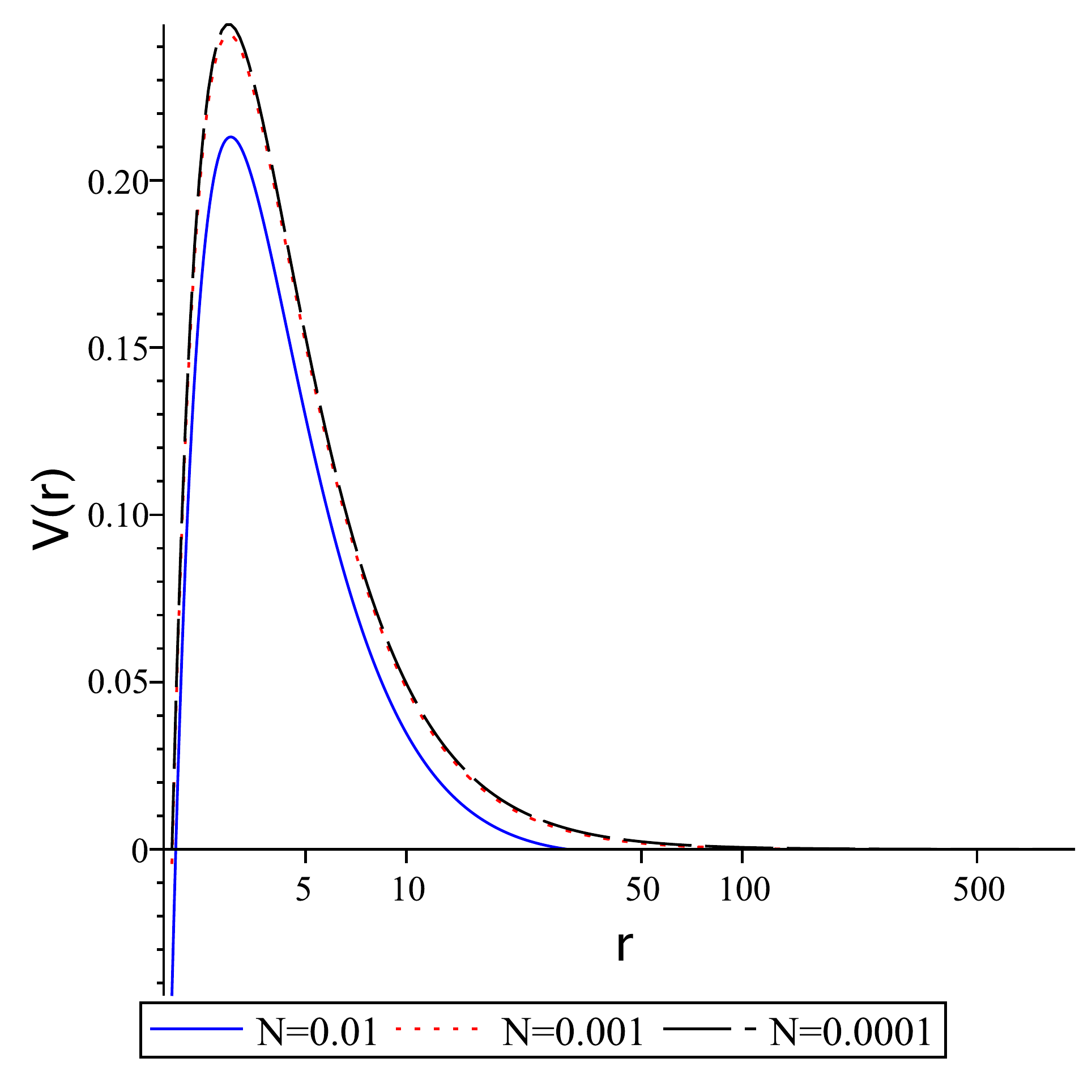} \\
\includegraphics[scale=0.4]{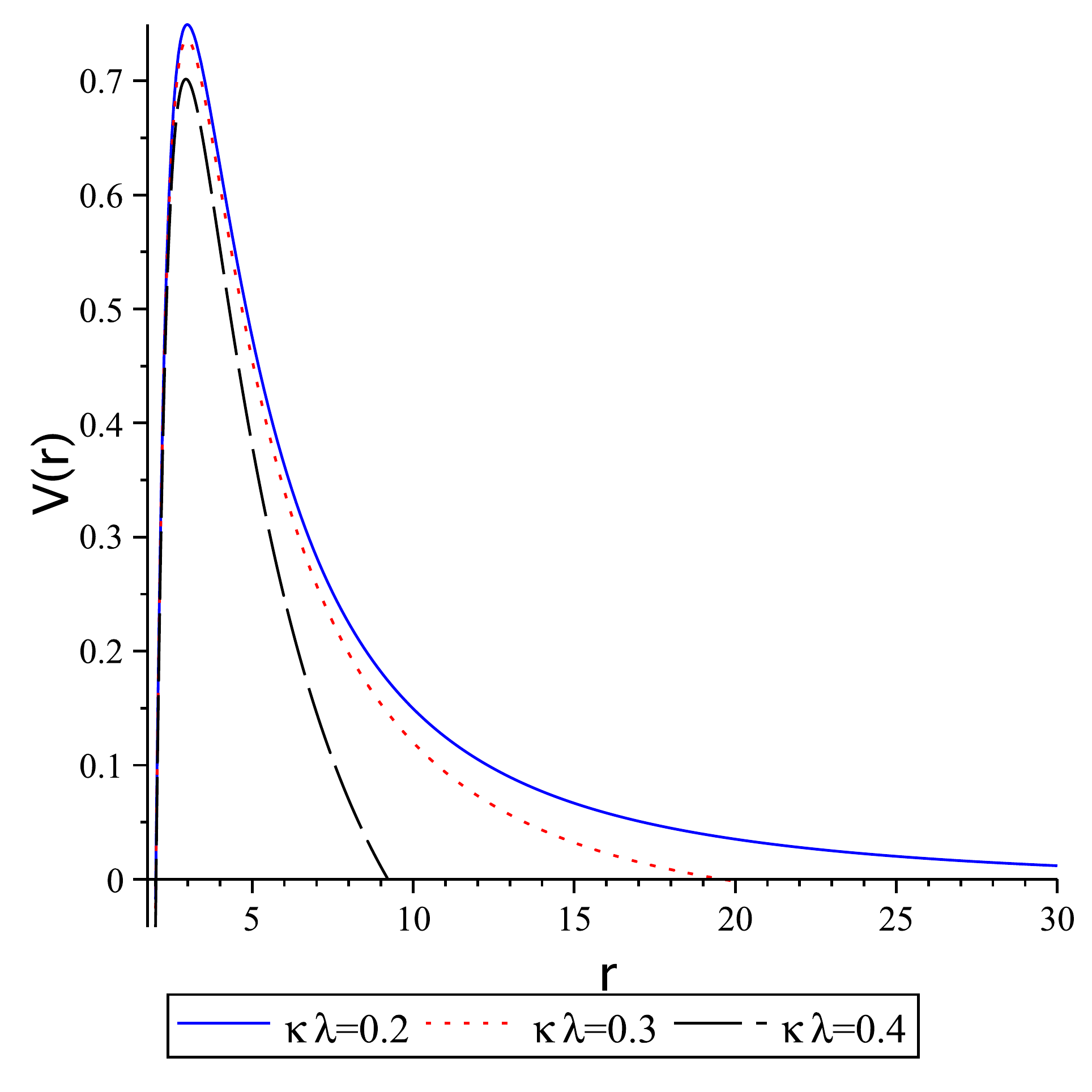} &
\includegraphics[scale=0.4]{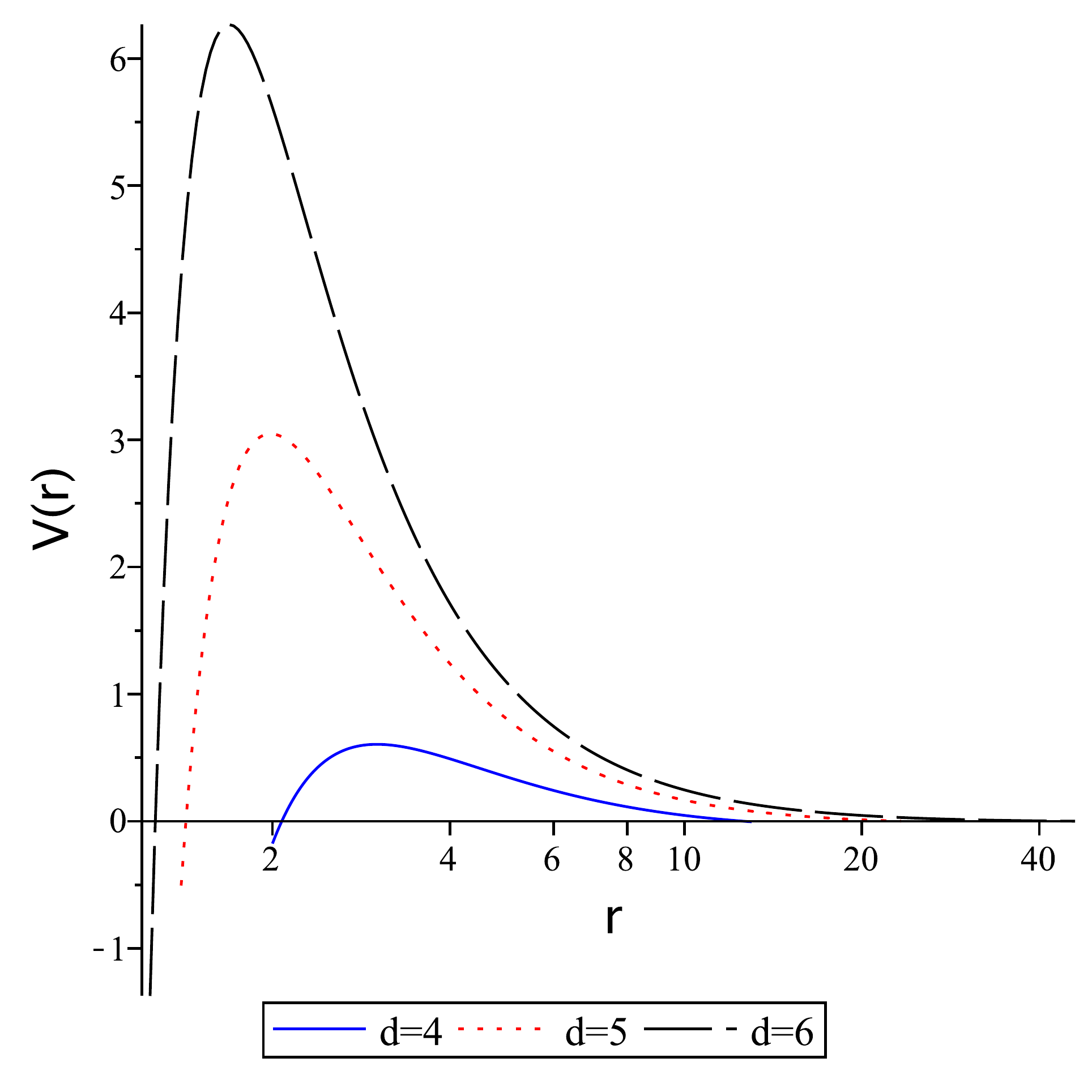} 
\end{tabular}
\caption{Upper left: $\omega=-2/3, \lambda \kappa = 0.1, M=1, N_s=0.01$ and $d=4$. Upper right: $\omega=-2/3, \lambda \kappa = 0.1, M=1, l=2$ and $d=4$. Bottom left: $\omega=-2/3, N_s=0.001, M=1, l=4$ and $d=4$. Bottom right: $\omega=-2/3, \lambda \kappa = 0.2, M=1, N_s=0.01$ and $l=4$}
\label{fig:Potential}
\end{figure}

In figure (\ref{fig:Potential}) it is plotted several configurations for the potential. Let us first note that due to the $f(r)$ term on the effective potential, it will vanish on both horizons. In the upper left figure, we vary the angular momentum eigenvalue using $\omega=-2/3, \lambda \kappa = 0.1, M=1, N_s=0.01$ and $d=4$. The result is similar the one obtained for the Schwarzschild metric. In the upper right figure, we vary the parameter $N_s$, related to the exotic matter field, using $\omega=-2/3, \lambda \kappa = 0.1, M=1, l=2$ and $d=4$. As we increase the parameter, the radius of the cosmological horizon decreases. For the bottom left figure, we vary the Rastall parameter $\kappa \lambda$ using $\omega=-2/3, N_s=0.001, M=1, l=4$ and $d=4$. One can note that the cosmological horizon decreases as we increase the Rastall parameter. For the bottom right figure, we vary the spacetime dimension, using $\omega=-2/3, \lambda \kappa = 0.2, M=1, N_s=0.01$ and $l=4$. One can note that the value of the cosmological parameter increases as we increase the spacetime dimension. It also can be noted that the event horizon decreases as the spacetime dimension increases, but this is a well-known feature of the Schwarzschild-Tangherlini metric.

To find the scalar quasinormal modes, one must solve equation (\ref{scalarEquation}), together with the following set of boundary conditions,

\begin{equation}
\Phi(\textbf{x} \rightarrow - \infty) = C_I \Phi_I^{out}, \hspace{10pt} \Phi(\textbf{x} \rightarrow + \infty) = C_{II} \Phi_{II}^{out}.
\label{boundaryConditions}
\end{equation}

Equation (\ref{scalarEquation}) can be solved using the familiar WKB approach, were an approximate WKB solution is found at approaching both the event and cosmological horizon, and matched with the approximated solution found around the peak of the potential. To conciliate the boundary conditions with the matching, the modes should obey the constraint

\begin{equation}
i\frac{Q_0}{\sqrt{2 Q_0''}} - \sum_{k=2}^{m} \Lambda_k = n + 1/2,
\end{equation}
where $Q(r) = E^2 - V(r)$, and $Q_0'' = d^2Q(r) / dr^{*2}$. The parameter $n$ is called the $tone$ number of the quasinormal mode.  The $\Lambda_k$'s can be obtained from the potential $Q(r)$ up to its $2k$-derivative, and indicates the order of the WKB method. The 3rd order WKB was developed in \cite{Iyer:1986np} and the 6th order in \cite{Konoplya:2003ii}. An explicit formula for $\Lambda_k$ can be found in the original papers \cite{Iyer:1986np}\cite{Konoplya:2003ii}. For $l > n$, the WKB method is a good approximation for the quasinormal modes, and in this paper we will use the 5th order WKB method to calculate them. Our aim is to determine the (quasinormal) modes $E$ of the system (we are not using $\omega$ to avoid confusion with the parameter $\omega$ of the equation of state).

\begin{figure}[]
\centering
\begin{tabular}{@{}cc@{}}
\includegraphics[scale=0.6]{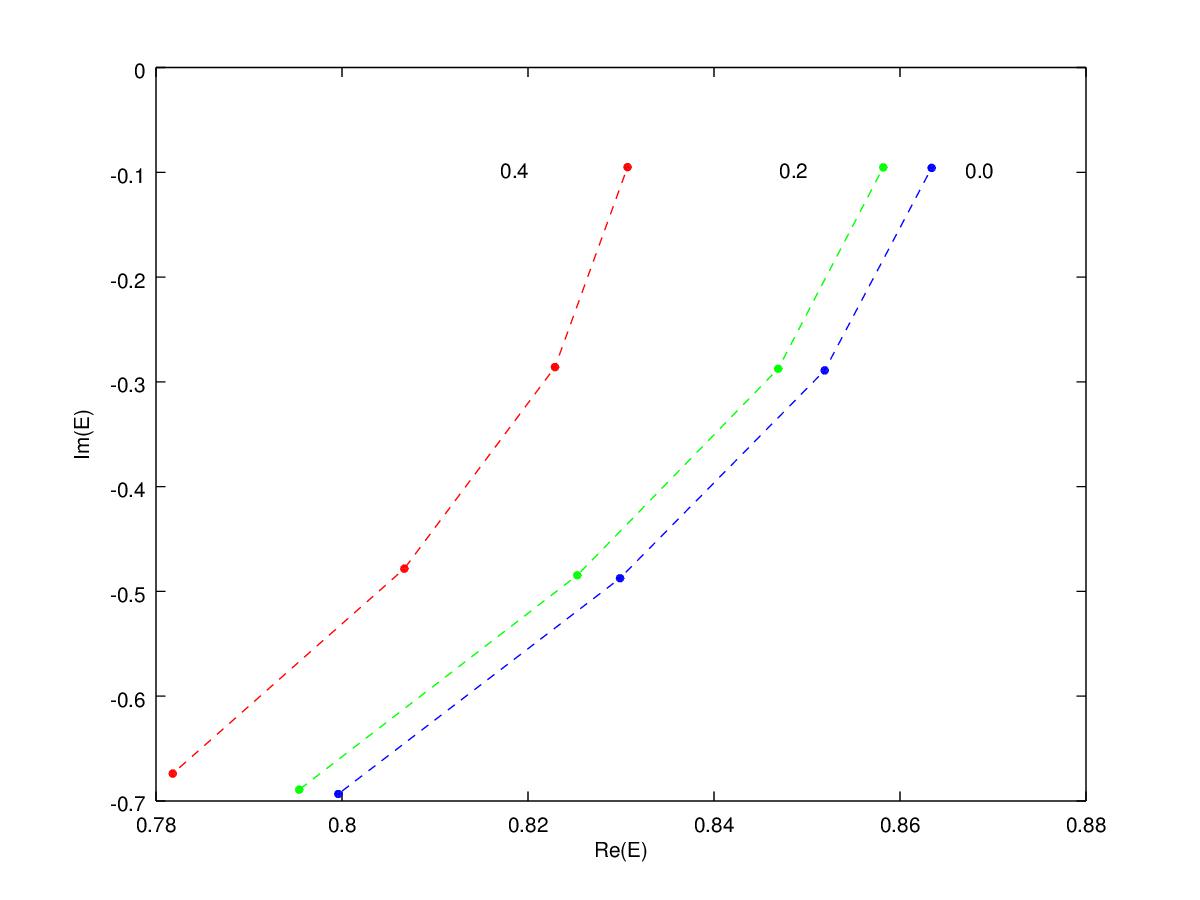} 
\end{tabular}
\caption{Scalar quasinormal modes for the parameters $M=1, N=0.001$ and $d=4$. We are using $l=4$, and the dots indicate, from top do bottom, $n=0,1,2$ and $3$. The blue, green and red lines were calculated for $\kappa \lambda=0.0, 0.2 $ and $0.4$.}
\label{fig:N001}
\end{figure}

\begin{figure}[]
\centering
\begin{tabular}{@{}cc@{}}
\includegraphics[scale=0.6]{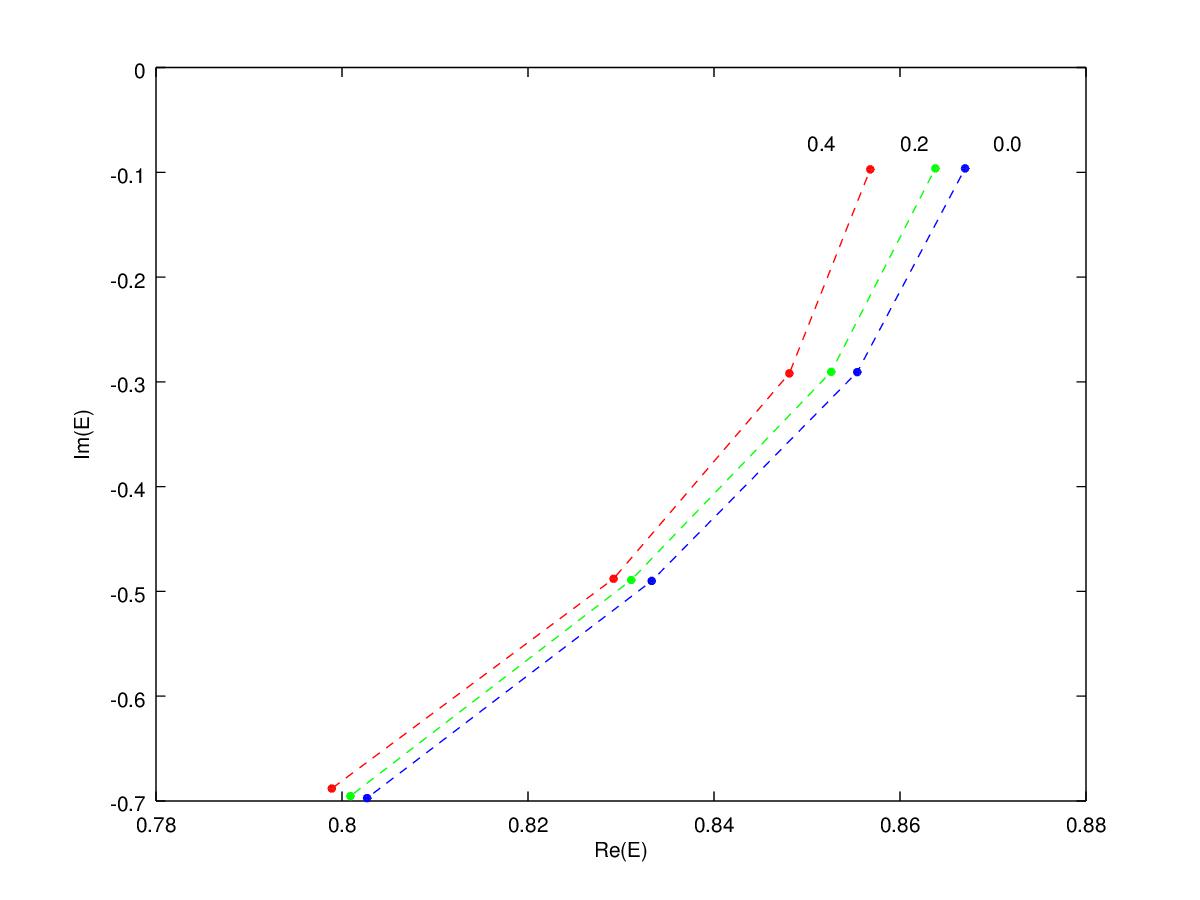} 
\end{tabular}
\caption{Scalar quasinormal modes for the parameters $M=1, N=0.0001$ and $d=4$. We are using $l=4$, and the dots indicate, from top do bottom, $n=0,1,2$ and $3$. The blue, green and red lines were calculated for $\kappa \lambda=0.0, 0.4 $ and $0.6$.}
\label{fig:N0001}
\end{figure}

As long this is a paper on Rastall gravity, we will study how the quasinormal modes change as we change the Rastall parameter $\kappa \lambda$. However, the model have a great number of free parameters, to known, the parameter $N_s$ related to the perfect fluid density, the equation of state $\omega$ and the dimension of the theory. We will start studying the effect of the Rastall parameter on the modes, in respect with a chosen parameter $N_s$. 

As can be seen in figure (\ref{fig:Potential}), at top right, the second zero of the potential (the cosmological horizon) is smaller for smaller values of the parameter $N_s$. On the same graph, at the left bottom, the cosmological horizon is smaller as we increase the Rastall parameter. This indicates that the quasinormal modes will be more affected for small values of the parameter $N_s$, coupled with larger values of the Rastall parameter. This behaviour can indeed be seen in figures (\ref{fig:N001}) and (\ref{fig:N0001}). In these figures, we keep the parameters $M, \omega$ and $d$ fixed, and vary the Rastall parameter for two values of the parameter $N_s$, to known, $N_s=0.001$ in figure (\ref{fig:N001}) and $N=0.0001$ in figure (\ref{fig:N0001}). When $N=0.001$, small values of the Rastall parameter induce a larger variation on the modes than greater values when $N=0.0001$. In fact, for larger values of the Rastall parameter than the ones plotted on these graphs, the event horizon is too close to the cosmological horizon, and so we cannot trust the results of the WKB method. If we increase the Rastall parameter a little bit more, there is no more horizons, and so there is no black hole.

\begin{table}[]
\centering
\caption{Shift on the real and imaginary parts of the QNMs, compared with the QNMs measured in GR ($\kappa \lambda \rightarrow 0$). The five columns are: dimension ($d$), used value for the parameter $N_s$, used value for the parameter $\kappa \lambda$, modulus the difference between the calculated quasinormal modes and the same correspondent modes on GR, both the real part and imaginary part. The others parameters were $l=4$, $n=0$ and $\omega=-2/3$.}
\label{tab:QNM23}
\begin{tabular}{p{0.15\textwidth}p{0.15\textwidth}p{0.15\textwidth}p{0.15\textwidth}p{0.15\textwidth}}
\hline
 $d$  & $N_s$  & $\kappa \lambda$ & $|\delta(E_{re})|$ & $|\delta(E_{im}|)$ \\ \hline
 5  & 0.001  & 0.6 & 0.0483 & 0.0118  \\ 
 5  & 0.0001  & 0.6 & 0.0047 & 0.0013  \\
 6  & 0.001  & 0.6 & 0.0236 & 0.0066  \\ 
 6  & 0.0001  & 0.6 & 0.0024 & 0.0007  \\
 7  & 0.001  & 0.6 & 0.0173 & 0.0052  \\ 
 7  & 0.0001  & 0.6 & 0.0017 & 0.0005  \\  
\end{tabular}
\end{table}

To push the discussion further, one can change the dimension of the theory or the equation of state. We will start changing the dimension of the theory, and look how the modes change as we change the Rastall parameter. We did the calculation for two values of the parameter $N_s$, and the behaviour of the modes is shown on table (\ref{tab:QNM23}). We are not so interested in the values of the modes, but the shift of the modes as related to the spacetime dimension. Table (\ref{tab:QNM23}) contains five columns: The dimension, the value of the parameter $N_s$, the value of the Rastall parameter and the shift of the mode between the value obtained with this Rastall parameter in relation with the one obtained in GR ($\kappa \lambda \rightarrow 0)$, for its real and imaginary parts. For the obtained values, we used $M=1, \omega=-2/3, l=4$ and $n=0$. 

As can be seen on table (\ref{tab:QNM23}), the shift on the modes is smaller as we decrease the parameter $N_s$ and increase the dimension of the spacetime. The result due the parameter $N_s$ was already clear on the relation between figures (\ref{fig:N001}) and (\ref{fig:N0001}), and the results due to the increasing on the spacetime dimension, although not obvious, is consistent with the potentials plotted on figure (\ref{fig:Potential}). There we can see that, as we increase the spacetime dimension, the cosmological horizon also increases. This indicates that the Rastall parameter, that is in someway related to the value of the cosmological horizon, will be less important for larger dimensions.

\begin{table}[]
\centering
\caption{Shift on the real and imaginary parts of the QNMs, compared with the QNMs measured in GR ($\kappa \lambda \rightarrow 0$). The five columns are: dimension ($d$), used value for the parameter $N_s$, used value for the parameter $\kappa \lambda$, modulus the difference between the calculated quasinormal modes and the same correspondent modes on GR, both the real part and imaginary part. The others parameters were $l=4$, $n=0$ and $\omega=-4/3$.}
\label{tab:QNM43}
\begin{tabular}{p{0.15\textwidth}p{0.15\textwidth}p{0.15\textwidth}p{0.15\textwidth}p{0.15\textwidth}}
\hline
 $d$  & $N$  & $\kappa \lambda$ & $|\delta(E_{re})|$ & $|\delta(E_{im})|$ \\ \hline
 4  & 0.001  & 1.0 & 0.0344 & 0.0009  \\ 
 4  & 0.0001  & 1.0 & 0.0033 & 0.0002  \\
 5  & 0.001  & 1.0 & 0.0166 & 0.0007  \\ 
 5  & 0.0001  & 1.0 & 0.0016 & 0.0001  \\
 6  & 0.001  & 1.0 & 0.0138 & 0.0003  \\ 
 6  & 0.0001  & 1.0 & 0.0014 & 0.0005  \\  
\end{tabular}
\end{table}

To study other kinds of exotic matter, one can change the equation of state, $\omega$. Here we will do the same analysis as sketched above for $\omega = -4/3$, a kind of field usually known as phantom field ($\omega < 1$). The calculated values are shown in table (\ref{tab:QNM43}), where, once more, we are interested on the shift of the modes relative to the values obtained in GR. We can note the same kind of behaviour found for the quintessence, but now we have been able to use values for the Rastall parameter close to unity, and could use even larger, without collapsing both horizons. The reason is that black hole solutions, with a cosmological horizon, exists for larger values of the Rastall parameter compared with the quintessence. As long as the Rastall parameter should be unique, the existence of black holes surrounded by a perfect fluid, in Rastall theory, should be a function of the parameter $N_s$, that depends on the density of the field configuration.

Our study could be enlarged to calculate the quasinormal modes for other kinds of exotic matter, but it is expected that the same behaviour will be realized. The main result of this study on black holes in higher dimension, surrounded by quintessence, in Rastall gravity, is that its existence imposes some bounds on the Rastall parameter, and this bound is strongly dependent on the density of the field, and the kind of exotic matter. For a phantom field, for example, the existence of black holes and the shift on the modes are less affected by the Rastall parameter than for a quintessence field, for the same energy density configuration.

\section{Conclusions}

In this paper, we found the spacetime generated by a black hole, surrounded by an exotic matter field, in Rastall gravity, for an arbitrary d-dimensional spacetime. Our main goal was to study the scalar quasinormal modes generated by perturbation on a probe field by such spacetime. 

Due to the large number of free parameters of the model, such as the equation of state of the exotic matter, the energy density of the field, the dimension of spacetime, and the Rastall parameter of the theory $\kappa \lambda$, we focus on the importance of the Rastall parameter for the existence of black holes, and the shift that such modification of the gravitational theory would impose on the quasinormal modes as calculated on the framework of general relativity, where $\kappa \lambda \rightarrow 0$.

We have shown that the existence of black holes strongly depends on the relation between all the free parameters of the theory. For a quintessence field ($\omega = -4/3$), the parameter $N_s$ must be small, otherwise the cosmological and event horizons will be too close to provide a real scenario for an astrophysical object. In this regime, the shift on the quasinormal modes are weak. The case of a phantom field allows for larger values on the parameter $N_s$ or for a fixed parameter $N_s$, allows for larges values of the Rastall parameter. 

When we increase the dimension of the spacetime, the effect of both the parameters $N_s$ and the Rastall parameter become weaker relative to the lower dimensional case. This means that, for the same set of parameters $N_s$ and $\kappa \lambda$, higher dimensions induce a lower shift on the quasinormal modes, compared with general relativity.

As long it is expected that the density of the exotic matter that supposedly pervades the universe is low, and the Rastall parameter should be of order unity, the measured shift on the quasinormal modes is not enough to allow some kind of instability on the system. To confirm such statement, one should study perturbations on the Rastall equations itself, and calculate the tensor quasinormal modes of the theory. We hope to proceed with this calculation in the near future.

%
%
\ack  IPL is supported by Conselho Nacional de Desenvolvimento Cient\'ifico e Tecnol\'ogico (CNPq-Brazil) by the grant No. 150384/2017-3. JPMG is supported by CAPES (Coordena\c{c}\~ao de Aperfei\c{c}oamento de N\'ivel Superior).

%
%

\end{document}